\documentclass[twocolumn,showpacs,preprintnumbers,amsmath,amssymb,superscriptaddress]{revtex4}

\usepackage{epsf}
\usepackage{graphicx}
\usepackage{sidecap}

\usepackage{color}

\def \beq {\begin{equation}}
\def \eeq {\end{equation}}
\begin{document}

\title{{Observation of Dirac-Like Semi-Metallic Phase in NdSb}}



\author{Madhab~Neupane}
\affiliation {Department of Physics, University of Central Florida, Orlando, Florida 32816, USA}

\author{M.~Mofazzel~Hosen}\affiliation {Department of Physics, University of Central Florida, Orlando, Florida 32816, USA}

\author{Ilya~Belopolski}
\affiliation {Joseph Henry Laboratory and Department of Physics, Princeton University, Princeton, New Jersey 08544, USA}


\author{Nicholas Wakeham}
\affiliation {Condensed Matter and Magnet Science Group, Los Alamos National Laboratory, Los Alamos, NM 87545, USA}

\author{Klauss~Dimitri}\affiliation {Department of Physics, University of Central Florida, Orlando, Florida 32816, USA}

\author{Nagendra Dhakal}
\affiliation {Department of Physics, University of Central Florida, Orlando, Florida 32816, USA}








\author{Jian-Xin~Zhu} \affiliation {Theoretical Division, Los Alamos National Laboratory, Los Alamos, NM 87545, USA}



\author{M.~Zahid~Hasan}
\affiliation {Joseph Henry Laboratory and Department of Physics,
Princeton University, Princeton, New Jersey 08544, USA}

\author{Eric D. Bauer}
\affiliation {Condensed Matter and Magnet Science Group, Los Alamos National Laboratory, Los Alamos, NM 87545, USA}

\author{Filip Ronning}
\affiliation {Condensed Matter and Magnet Science Group, Los Alamos National Laboratory, Los Alamos, NM 87545, USA}

\date{18 June, 2013}
\begin{abstract}

{The search of new topological phases of matter is one of the new directions in condensed matter physics. Recent experimental realizations of Dirac semimetal phases pave the way to look for other exotic phases of matter in real materials. Here we present a systematic angle-resolved photoemission spectroscopy (ARPES) study of NdSb, a potential candidate for hosting a Dirac semi-metal phase. 
Our studies reveal two hole-like Fermi surface pockets present at the zone center ($\Gamma$) point as well as two elliptical electron-pockets present in the zone corner (X) point of the Brillouin zone (BZ).
Interestingly, Dirac-like linearly dispersive states are observed about the zone corner (X) point in NdSb. Our first principles calculations agree with the experimentally observed bands at the $\Gamma$ point.  Moreover, the Dirac-like state observed in NdSb may be a novel correlated state, not yet predicted in calculations. Our study opens a new direction to look for Dirac semi-metal states in other members of the rare earth monopnictide family.}

\end{abstract}
\date{\today}
\maketitle

The discovery of topological Dirac and Weyl materials have gained intense research interest both theoretically and experimentally as they can host many different novel phenomena such as linear bulk band crossings, surface Fermi arcs, chiral anomalies, perfect compensation between electron and hole carriers, resistivity plateaus, Shubnikov-de Hass (SdH) oscillations, etc \cite{Hasan, SCZhang, Hasan_review_2, Xia, Neupane, Neupane_2, Neupane_1, Nagaosa, Suyang_Science, Hong_Ding, TaAs_theory_1, TaAs_theory, Neupane_3}. 
These semi-metals (not insulators) show that the conduction
and valence bands disperse linearly through the crossing points in all directions throughout the three-dimensional momentum space. Bands are doubly degenerate in Dirac semi-metals, whereas the degeneracy is lifted by breaking either the time-reversal or inversion symmetry in Weyl semi-metals.
The linear band crossing at the Dirac point in a Dirac semi-metal is protected by the crystalline symmetry. Interestingly, extreme magnetoresistance (XMR) was found in Dirac semi-metals such as in Na$_3$Bi and Cd$_3$As$_2$ \cite{Ong_1, Ong_2}, which is further extended to Weyl semi-metals such as the TaAs family \cite{TaAs_1} and layered semi-metals such as WTe$_2$ \cite{Cava_1}. XMR materials are observed to be semi-metals with nearly compensated electron and hole carriers. Recently, XMR has also been reported in rare earth monopnictides such as LaSb and LaBi \cite{Cava_2, Cava_3, LaBi_1, LaSb_1, LaBi_2}. 

Rare earth monopnictides possess the simple NaCl-type structure (see Figure 1(a)). Using first principles calculations, the topological semi-metal phases have been predicted in rare earth monopnictides with the presence of inversion symmetry \cite{Liang_Fu}. Due to the possibility of a topological Dirac semi-metal phase in rare earth monopnictides, research interest on these materials is growing. However, a detailed experimental electronic structure of these materials has not yet been reported.  
As a momentum-resolved probe of electronic structure, which can isolate the surface from bulk states, angle resolved photoemission spectroscopy (ARPES) can provide convincing evidence of topological Dirac semi-metal phases.
We choose neodymium antimonide (NdSb) for our electronic structure measurement, 
which exhibits unique structural, electronic, magnetic and phonon properties \cite{NdSb_1, NdSb_2}. Furthermore, the presence of the $f$-electrons in NdSb may open a new possibility of revealing correlated topological phases which are absent in conventional topological insulators and graphene.

Here we report the experimental observation of a Dirac-like semi-metal phase in NdSb using ARPES.  Our wider Brillouin zone (BZ) mapping reveals multiple pockets at the Fermi level. For example, we observe two hole-like Fermi surfaces at the zone center ($\Gamma$) point and interconnecting elliptical-shaped Fermi surfaces at  the zone corner X point. Interestingly, the states around the zone corner are linearly dispersive Dirac-like in nature. 
The Dirac-like states observed here may be influenced by correlation effects due to the presence of the $f$-electrons in NdSb system, which requires more theoretical investigation.
Our findings suggest that rare earth monopnictides with XMR can serve as a new platform to reveal Dirac-like phases.

\begin{figure*}
\centering
\includegraphics[width=17.2cm]{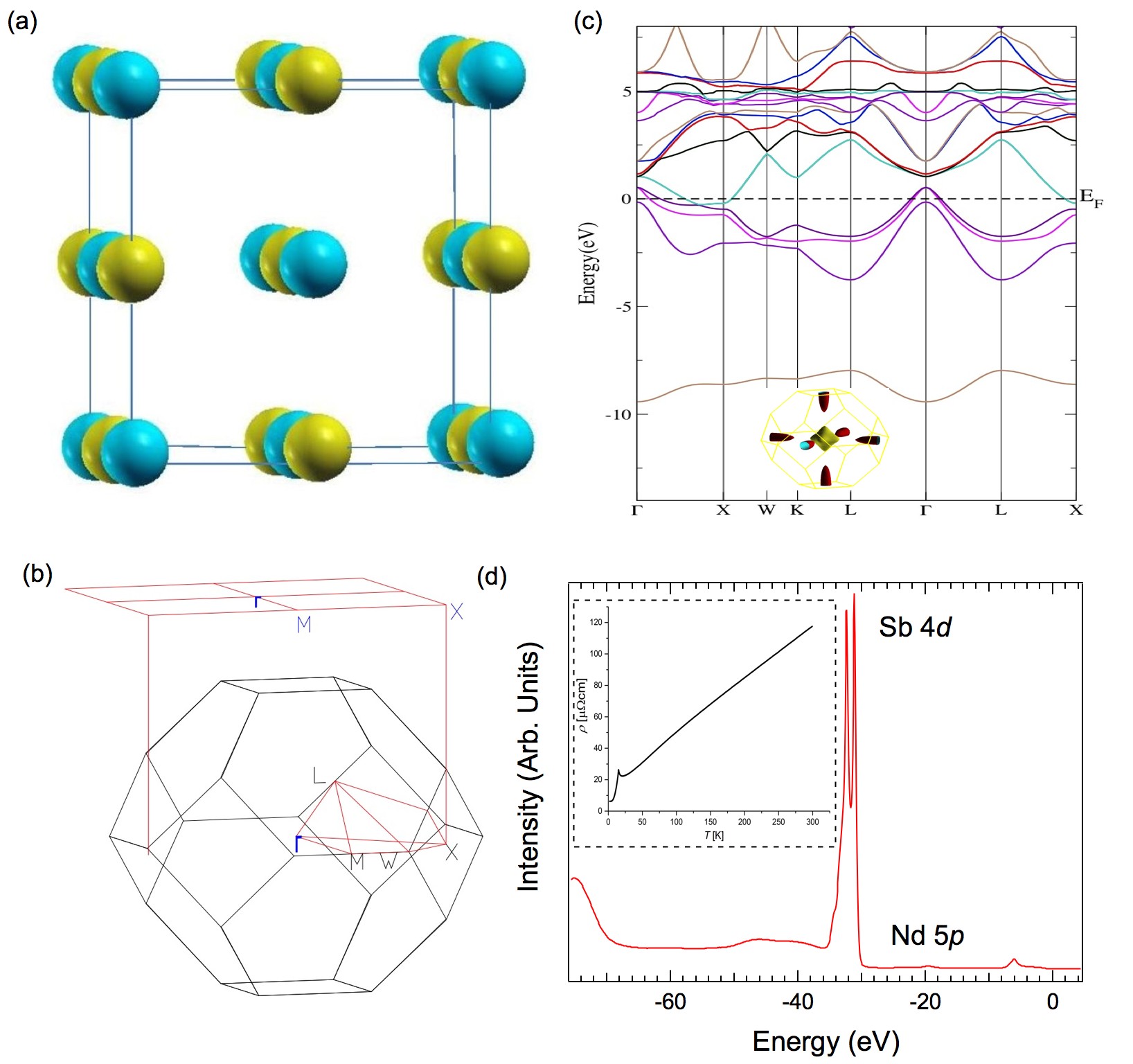}
\caption{{Brillouin zone, electronic structure and sample characterization of NdSb.}
(a) Rock-salt type crystal structure of NdSb, where yellow spheres represent Nd and blue spheres represent Sb. (b) Schematic 3D bulk Brillouin zone (BZ) and projected surface BZ of face-centered cubic lattice NdSb. High symmetry points are also marked. (c) First-principles band structure calculations along the high symmetry lines of bulk NdSb. Fermi level is also shown by dashed line. Inset shows the calculated bulk Fermi surface. (d) Core level spectrum of NdSb, which shows the sharp peak of Sb \textit{4d} and Nd \textit{5p}. This data is collected with a photon energy of 100eV. Inset shows the resistivity of NdSb as a function of temperature. A peak at $\sim$ 15 K indicates the magnetic transition temperature (T$_N$). 
Photoemission spectroscopy (PES) data were collected at SIS-HRPES end-station at SLS, PSI.}
\end{figure*}

Single crystals of NdSb were grown by the Sn flux technique as described elsewhere \cite{Material}. 
Synchrotron-based ARPES measurements of the electronic structure  were performed at the SIS-HRPES end-station of the Swiss Light Source and ALS BL 10.0.1 with
a Scienta R4000 hemispherical electron analyzer. The energy resolution was set to be better
than 20 meV. The angular resolution was set to be better than  0.2$^{\circ}$ for the synchrotron measurements. 
First principles density functional theory calculations were performed to compute the electronic structure using the WIEN2K code. The PBE functional under the generalized gradient approximation was employed and spin orbit coupling was included via a second order variational scheme and the three Nd $f$-electrons were treated as core electrons \cite{GGA_1}. NdSb was computed using the rock salt structure with lattice parameters of $a$ = 6.319 \AA ~ \cite{cal_2}.


\begin{figure*}
\centering
\includegraphics[width=17.50cm]{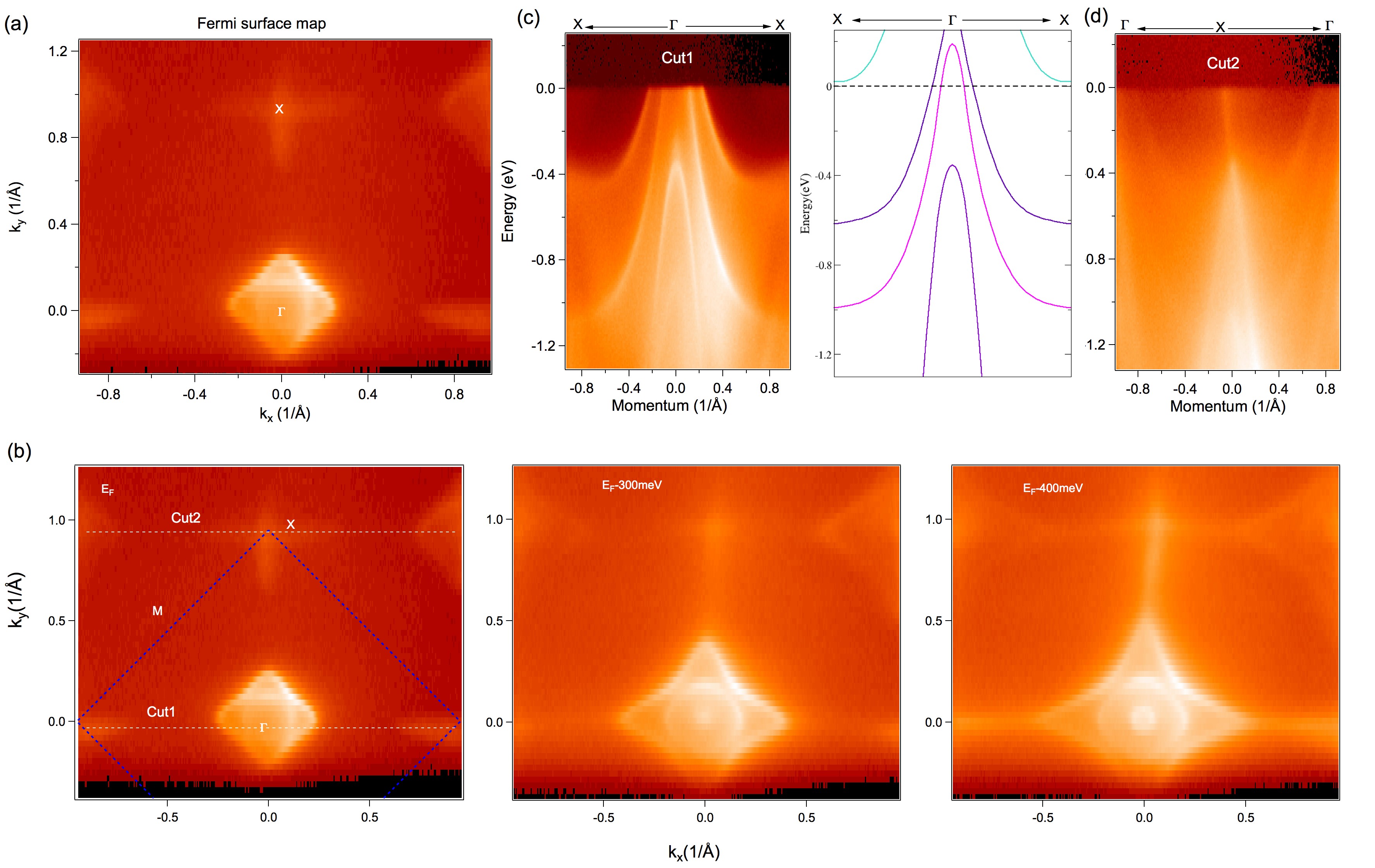}
\caption{{Measured electronic dispersion of NdSb.} (a) ARPES measured Fermi surface mapping along the X-$\Gamma$-X direction with incident photon energy 50eV at a temperature of 20 K. (b) Fermi surface and constant energy contour map along the X-$\Gamma$-X direction. Blue dash line indicates the surface Brillouin Zone. White dash lines noted as cut1 and cut2 are the momentum directions of the dispersion maps shown in (c) and (d). The binding energies are noted on the constant energy contour plots. (c) Dispersion map along the X-$\Gamma$-X direction as indicated in (b) (left). The right figure shows the calculated band structure with k$_z$ = 0.1$\pi$/a. (d) Dispersion map along the $\Gamma$-X-$\Gamma$ direction as indicated in (b) (left). 
ARPES data were collected at SIS-HRPES end-station at SLS, PSI with a photon energy of 50eV and temperature of 20 K}.
\end{figure*}

\begin{figure*}
\centering
\includegraphics[width=12.0cm]{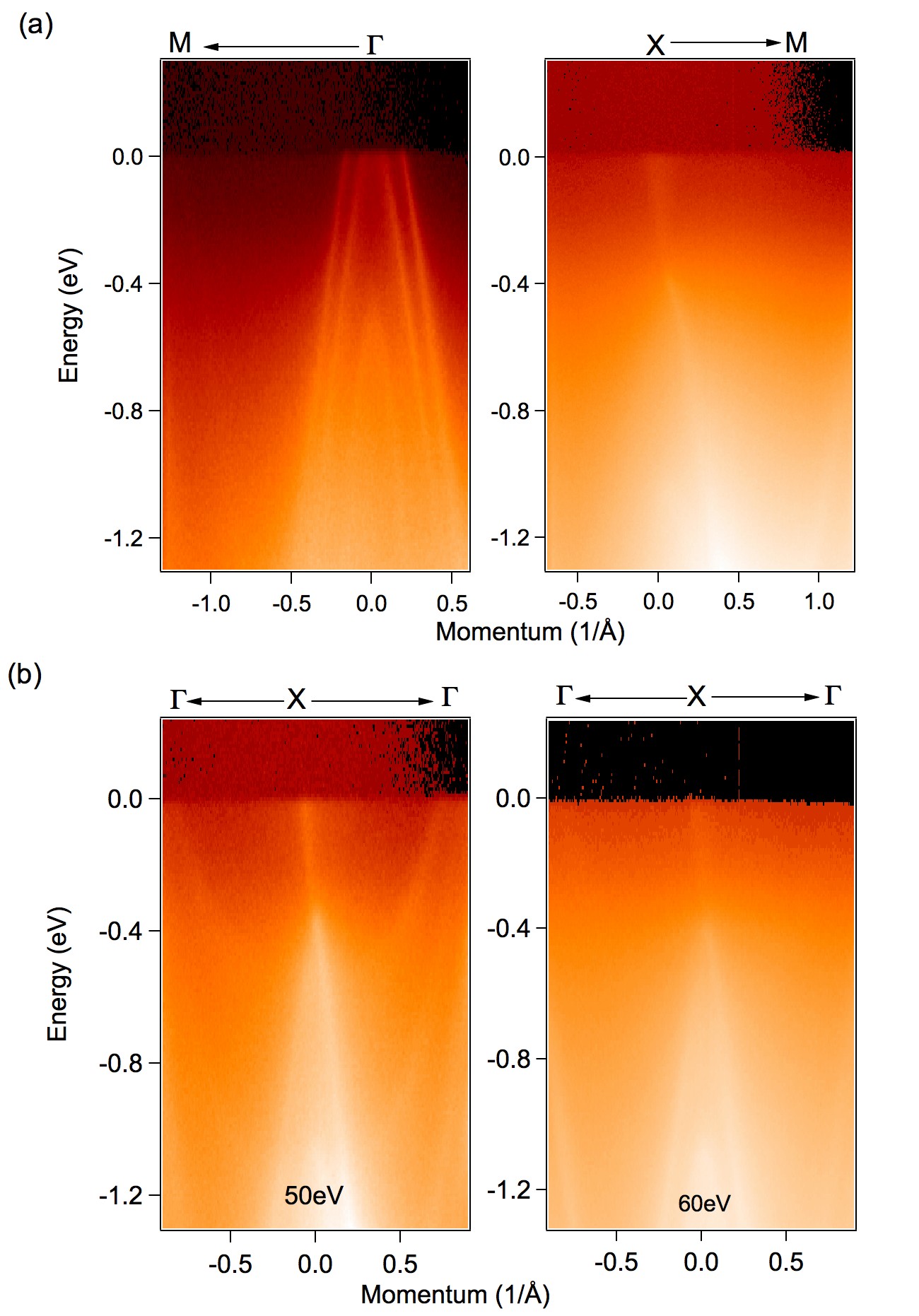}
\caption{\textbf{Dispersion maps along various high-symmetry directions.} 
(a) Dispersion map measured along the M-$\Gamma$-M and M-X-M direction. (b) Dispersion maps along the $\Gamma$-X-$\Gamma$ momentum direction measured with photon energy of 50eV and 60eV. Both measurements show the linear dispersive bands at around X point.}
\end{figure*}

We start our discussion by presenting the crystal structure of NdSb. It crystallizes in a NaCl-type crystal structure with space group $Fm\bar{3m}$ (see Figure 1(a)). 
Figure 1(b) shows the bulk and surface Brillouin zone (BZ) of NdSb. The high-symmetry points are marked on the bulk BZ. The center of the surface BZ is the $\Gamma$ and the corner of the BZ is the X point as indicated in Figure 1(b). 
The calculated bulk band structure along the high-symmetry direction of the Brillouin zone is shown in Figure 1(c). Two hole-like bands cross the Fermi level while a third band is about 0.4eV below the Fermi level at the $\Gamma$ point. An electron-like band also crosses the Fermi level at the X point. Inset of Figure 1(c) shows the Fermi surface plot obtained from bulk band calculations of NdSb. 
Photoemission spectroscopy also provides the core level of the orbital relative to the chemical potential. 
Figure 1(d) shows the core levels in the 0-70eV binding energy range of NdSb. We observe the Sb 4$d$ ($\sim$ 33eV) and Nd 5$p$ ($\sim$ 20eV) states. 
We observe a small peak about $\sim$ 6eV below the Fermi level, which is mostly coming from Nd 4$f$-orbitals. This justifies our treatment of the Nd $f$-electrons as core electrons in the calculations.
The observation of sharp peaks suggests that the samples used in our spectroscopic measurements are of high quality. 
The inset in Figure 1(d) shows resistivity in zero field up to room temperature, which demonstrates a clear peak in the resistivity at the Neel temperature (T$_N$) $\sim$ 15 K. XMR of 1.2 $\times$ 10$^4$$\%$ at 2 K and 9T has also been observed in NdSb \cite{Filip}.


In order to determine the detailed electronic structure of NdSb, we present the results of our systematic electronic structure studies in Figures 2 and 3. Figure 2(a) shows the plot of the Fermi surface map measured experimentally. We observe various pockets at the Fermi level. 
Figure 2(b) shows the plot of the Fermi surface map and constant energy contours measured with various binding energy noted in the plots, which is measured using a photon energy of 50eV. 
The blue dashed line shows the Brillouin zone and the white dashed lines marking cut1 and cut2 indicate the direction of the dispersion maps presented in Figures 2(c)-(d).
Experimentally, we observe two circular-shaped Fermi surfaces at the zone center and two overlapping elliptical Fermi pockets are found at the zone corner.  At 300 meV below the Fermi level, the pockets around the zone center grow in size, thus confirming the hole-like nature of the bands, where the size of the bands at the zone corner decrease upon increasing the binding energy confirming their electron-like nature.
Moving towards even higher binding energies, the pockets around the $\Gamma$ point further increase in size, Figure 2(b) (right). 
Figure 2(c) (left) and 2(d) show the measured dispersion maps along the high symmetry direction X-$\Gamma$-X and $\Gamma$-X-$\Gamma$, respectively. Two hole-like bands crossing the Fermi level are clearly observed at the zone center point. Another hole-like band is found to be located about 0.4eV below the Fermi level. Experimentally observed bands at the zone center agree well with the bulk band calculations. The optimal agreement was found using k$_z$ = 0.1$\pi$/a (see Figure 2(c) (right)).  Interestingly, a linear Dirac-like dispersive band is observed at the X point, with the energy location of the Dirac point at about 370 meV from the Fermi level. Importantly, such a linear Dirac-like dispersive band observed in NdSb at the X-point is not present in first principles calculations of the bulk (see Figure 2(c) right). Our experimental observations demand further theoretical calculations on this system in order to understand the Dirac semi-metallic states observed experimentally. Moreover, such a Dirac like semi-metal phase is theoretically predicted in non-$f$ electron system such as LaSb and LaBi \cite{Liang_Fu}.

In order to better understand the electronic structure of NdSb, we present dispersion maps of NdSb along different high symmetry directions. Figure 3(a) shows the dispersion maps measured along the M-$\Gamma$-M (left) and the M-X-M (right) momentum space cuts. These plots reveal the hole like bands at the $\Gamma$ point and linearly dispersive band at the X point, which is consistent with the data presented in Figure 2. Moreover, the dispersion maps measured along the $\Gamma$-X-$\Gamma$ direction with a photon energy of 50eV and 60eV are shown in Figure 3(b). Both of these spectra show the linearly dispersive bands in the vicinity of the X-point, thus confirming the presence of the Dirac-like state in NdSb.

Now we discuss a few observations of our measurements. First, our experimental data do not show the Nd 4$f$- band near the Fermi level. 
Our core level measurements show the Nd 4$f$-orbital at $\sim$ 6eV below the Fermi level. Second, the overall measured electronic structure of this system resembles iron-based superconductors  with multiple hole pockets at the zone center and two electron pockets at the zone corner as observed in 122 iron pnictide system \cite{Neupane_4}. 
Third, our experimental results negate the possibility of the possible topological insulator states in this system. There are many bands in the vicinity of the Fermi level, which suggests that NdSb is a semi-metal (metal) rather than an insulator. 
However, NdSb is topological in a similar sense. A Dirac-like surface state at the X point is expected due to a band inversion in the bulk band of the material \cite{Liang_Fu}. It is interesting that the surface state is robust even in the presence of bulk metallic states.
Fourth, it is reported that the magnetic transition temperature (T$_N$) of NdSb is $\sim$ 15 K (see Figure 1(d) inset). Our measured temperature is slightly higher than T$_N$. We expect that the overall electronic structure does not change below T$_N$ aside from zone folding which is supported by low temperature quantum oscillation measurements \cite{Filip}. 
Fifth, we would like to stress that the experimentally observed bands at the zone center are in good agreement with the first principles calculations (see Figure 2(c)). However, the bands at the zone corner are not reproducible from the calculations. Furthermore, we speculate that these Dirac-like bands are modified by correlation effects due to the presence of $f$-electrons in NdSb. Our experimental observations demand a detailed theoretical investigation of this system in order to reveal the interconnection of topology and correlation.   
 
 In conclusion, we perform systematic ARPES measurements on NdSb single crystals covering the entire area of the Brillouin zone. We reveal the existence of multiple Fermi pockets, which is complemented by the calculations. Interestingly, we find the linearly dispersive Dirac-like state at the zone corner X point, which is not present in the first principles band calculations and demands a new theoretical understanding of topology and correlation. Our study provides a new platform to search for Dirac semi-metal phases with XMR.

\bigskip
\bigskip
\bigskip
\hspace{0.5cm}
\textbf{Acknowledgements}
\newline

MN is supported by the start-up fund from University of Central Florida (UCF).
IB acknowledges the support of the NSF GRFP.
Work at Princeton University are supported by the Emergent Phenomena in Quantum Systems Initiative of the Gordon and Betty Moore Foundation under Grant No. GBMF4547 (MZH) and by the National Science Foundation, Division of Materials Research, under Grants No. NSF-DMR-1507585 and No. NSF-DMR-1006492. 
Synthesis and transport measurements of the crystals were performed under the auspices of the U.S. Department of Energy, Office of Science. Electronic structure calculations were performed with the support of the Los Alamos National Laboratory LDRD program.
We thank Plumb Nicholas Clark for beamline assistance at the SLS, PSI. We also thank Sung-Kwan Mo for beamline assistance at the LBNL. MN acknowledges the fruitful discussion with 
T Durakiewicz and A V Balatsky at LANL, and Talat Rahman at UCF.








\*Correspondence and requests for materials should be addressed to M.N. (Email: Madhab.Neupane@ucf.edu).

\end{document}